# Multi-track reconstruction algorithm in the Mu2e experiment

**Alessandro Maria Ricci**[a,*] **on behalf of the Mu2e Collaboration**

[a]*University of Pisa,*
 *Largo B. Pontecorvo, 3, I-56127, Pisa, Italy*

[a]*INFN Section of Pisa,*
 *Largo B. Pontecorvo, 3, I-56127, Pisa, Italy*

 *E-mail:* alessandro.ricci@df.unipi.it

The Mu2e experiment, under construction at Fermilab, will search for the neutrino-less coherent $\mu^- N \to e^- N$ conversion in the field of a $^{27}$Al nucleus. Such a process violates lepton flavor conservation. About 60% of muons stopped by an $^{27}$Al nucleus will undergo nuclear capture, while about 40% will decay in orbit. To quantify the conversion probability, we define $R_{\mu e}$, which is given by the ratio between the $\mu^- \to e^-$ conversion rate and the nuclear capture rate [1]:

$$R_{\mu e} = \frac{\Gamma\left(\mu^- + N\left(Z, A\right) \to e^- + N\left(Z, A\right)\right)}{\Gamma\left(\mu^- + N\left(Z, A\right) \to \nu_\mu^- + N\left(Z-1, A\right)\right)} \ . \tag{1}$$

The upper limit on $R_{\mu e}$ is $7 \cdot 10^{-13}$ at 90% CL, set by the SINDRUM II experiment [2]. The goal of the Mu2e experiment is to reach a sensitivity on $R_{\mu e}$ of $8 \cdot 10^{-17}$ at 90% CL. This represents a four-order of magnitude improvement over the current experimental limit.

Mu2e will take its first data in 2027. The signature for the muon conversion is a monochromatic electron of 104.97 MeV/c, an energy slightly below the muon rest mass. While the main experiment goal is to reconstruct the conversion electron, i.e., an event with a single track, there are motivations to develop an efficient tracking algorithm for reconstructing more simultaneous tracks. This could better constrain the background generated by $p\bar{p}$-annihilation in the Al target and to search for other Beyond the Standard Model processes. In this paper, we present an algorithm designed to reconstruct multi-particle events.

*Speaker

## 1. Track reconstruction algorithms

The main Mu2e detector is a straw-tube tracker in a uniform 1 T magnetic field [1]. It is an array of straw drift tubes perpendicular to the beam axis (z-axis). It is designed to accurately measure the trajectory of the electrons in order to determine their momenta.

The track reconstruction is divided in four sequential stages: 1) Hit Reconstruction: raw current signals are converted into position and time coordinates. 2) Time Clustering: hits close in time to each other are grouped together to create time clusters. 3) Helix Finding: within each time cluster, hits consistent with a helix are grouped into helix seeds. 4) Track Fit: the helix seeds are processed by a Kalman filter fit.

The default Mu2e algorithms form time clusters using an ANN trained to efficiently search for a conversion $e^-$, which removes a large fraction of pion and muon hits. We have developed new clustering algorithms, without any ANN, that are highly efficient for a wide spectrum of particles expected in multi-track events.

Assuming that the straw hits are produced by a particle with a constant momentum component $p_z$, the time clustering algorithm searches for straight lines in time vs z coordinate space, where z is the axis of the tracker. However, the so-obtained time clusters are inhomogeneous and contain hits of different particles. In many cases, they do not result in a reconstructed track. In some cases, the tracks are effectively simultaneous and the time information alone is not able to disentangle them (Fig. 1).

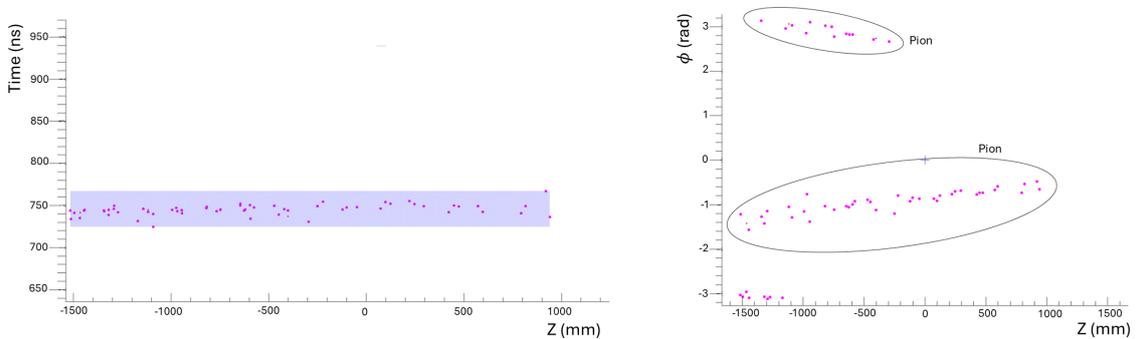

**Figure 1:** An example of hit separation in $\phi$ vs z coordinate space. The hits coming from two pions produced from a $p\bar{p}$-annihilation could not be adequately distinguished in time vs z plane (left), but are well-separable in $\phi$ vs z plane (right), where $\phi$ is the azimuth angle.

The hits produced by different particles with the same initial time and longitudinal speed overlap in the t-z plane but can look well separated in the $\phi$-z plane if they correspond to tracks with different angular velocity or a different offset $\phi_0$ (Fig. 1). Hits are associated to the same cluster if they belong to the same straight line in both the t-z and the $\phi$-z planes. These clusters have a higher purity and when they are used to feed the track reconstruction algorithm the efficiency and the quality of the track reconstruction improves. Fig. 2 shows the same event of $p\bar{p}$-annihilation in presence of the pile-up hits produced by other beam particles. The standard clustering algorithm finds the leading track, the new algorithm finds both real tracks.

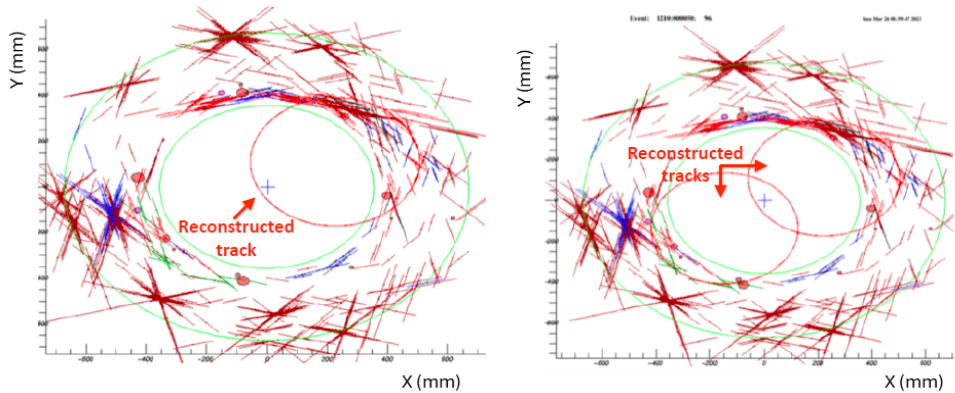

**Figure 2:** Example of track reconstruction of the same $p\bar{p}$-annihilation event in presence of the pile-up hits produced by other beam particles. The green circles are the tracker edges and the colored bars are straw hits. Left: standard clustering algorithm. Right: new clustering algorithm.

## 2. Results

When the multi-track cluster algorithm is applied to simulated $\mu^- \to e^-$ events with beam pile-up, it selects the 99.2% of time clusters containing the $\mu^- \to e^-$ process. When applied to events produced by $p\bar{p}$-annihilation in the stopping target superimposed with beam pile-up, it selects 90% of the time clusters containing $p\bar{p}$-annihilation products. It rejects 77% of the time clusters that do not contain these processes.

## 3. Acknowledgements

We are grateful for the vital contributions of the Fermilab staff and the technical staff of the participating institutions. This work was supported by the US Department of Energy; the Istituto Nazionale di Fisica Nucleare, Italy; the Science and Technology Facilities Council, UK; the Ministry of Education and Science, Russian Federation; the National Science Foundation, USA; the National Science Foundation, China; the Helmholtz Association, Germany; and the EU Horizon 2020 Research and Innovation Program under the Marie Sklodowska-Curie Grant Agreement Nos. 734303, 822185, 858199, 101003460, and 101006726. This document was prepared by members of the Mu2e Collaboration using the resources of the Fermi National Accelerator Laboratory (Fermilab), a U.S. Department of Energy, Office of Science, HEP User Facility. Fermilab is managed by Fermi Research Alliance, LLC (FRA), acting under Contract No. DE-AC02-07CH11359.